\documentclass[
superscriptaddress,
preprint,
nofootinbib,
amsmath,amssymb,
aps,
prd,
floatfix,
]{revtex4-2}
\usepackage{graphicx}
\usepackage{bm}
\usepackage[colorlinks=true,hyperfootnotes=true,breaklinks=true,citecolor=blue]{hyperref}
\usepackage{physics}
\begin{document}
\count\footins = 1000
\newcommand{\tbf}[1]{\textbf{#1}}
\def\bkt#1{\left(#1\right)}

\title {Foundational Issues in Dynamical Casimir Effect and Analogue Features in Cosmological Particle Creation}
\author{Jen-Tsung Hsiang}
\email{cosmology@mail.ntust.edu.tw}
\affiliation{College of Electrical Engineering and Computer Science, National University of Science and Technology, Taipei City, Taiwan 106, ROC}
\author{Bei-Lok Hu}
\email{blhu@umd.edu}
\affiliation{Maryland Center for Fundamental Physics and Joint Quantum Institute, University of Maryland, College Park, Maryland 20742, USA} 

\begin{abstract} 
Moving mirrors as analogue sources of Hawking radiation from black holes have been explored extensively, less so with cosmological particle creation (CPC), even though the analogy between dynamical Casimir effect (DCE) and  CPC based on the mechanism of parametric amplification of quantum field fluctuations has also been known for a long time. This `perspective' essay intends to convey some of the rigor and thoroughness of quantum field theory in curved spacetime, which serves as the theoretical foundation of CPC, to DCE, which enjoys a variety of active experimental explorations.  We have selected out seven issues of relevance to address, starting from the naively simple ones, e.g., why should one be bothered with `curved' spacetime when performing a laboratory experiment in ostensibly flat space, to foundational theoretical ones, such as the frequent appearance of nonlocal dissipation in the system dynamics induced by colored noises in its field environment, the existence of quantum Lenz law and fluctuation-dissipation relations  in the backreaction effects of  DCE emission on the moving atom/mirror or the source,  and the construction of a microphysics model to account for the dynamical responses of a mirror  or medium.  The strengthening of theoretical ground for DCE is useful not only for improving conceptual clarity but needed for the development of proof of concept type of future experimental designs for DCE. Results from DCE experiments in turn will enrich our understanding of quantum field effects in the early universe because they are, in the spirit of analogue gravity, our best hopes for the verification of these fundamental processes. 
\end{abstract}


\maketitle
\parskip=10pt


\section{Introduction} 
In this ``viewpoint/perspective" paper we  re-examine some basic  issues in dynamical Casimir effect (DCE) from minor (mostly known and correctly dealt with, yet worth pointing out as some popular papers have missed out) to moderate (technically sound,  yet conceptually not understood deeply or thoroughly enough) to major (may not have been taken into account fully, yet important for future explorations) using the more  developed and advanced  theoretical framework and better conceptual understanding gained from  the research on cosmological particle creation (CPC) since the late 60s to today \cite{BirDav,ParTom,HuVer}. 
In parallel to the 50 years of theoretical developments for CPC, some key issues of which are highlighted here, on the experimental side, a great variety of  DCE experiments have been carried out, or are actively in preparation. We are fortunate to have excellent reviews such as \cite{Dodonov} on DCE and \cite{AnalogG} for analogue gravity which keep up with the latest developments in both fields. The strengthening of theoretical ground for DCE is useful not only for a better conceptual clarity but needed for the development of proof of concept type of future experimental designs for DCE, which is our best hope for the verification of early universe quantum field processes in the spirit of analogue gravity.


\subsection{Three kinds of quantum field processes}
Both CPC and DCE encompass \textbf{three kinds of quantum field processes}, see Table 1 of \cite{XHH}:  1) {\it Casimir effect} \cite{Casimir}, the static limit of DCE, corresponding to a universe with a fixed and constant scale factor, such as the Einstein universe \cite{Ford75}, which by itself is of great interest and importance;  2) {\it particle creation} \cite{Par69,Zel70}, time varying scale factor (dynamical, nonadiabatic process) and  3) \textit{\it trace anomaly}.  Trace anomaly \cite{TraceA} (for its gravity context, see discussions in \cite{BirDav,ParTom,HuVer}) involves higher derivatives of the scale factor, which can play an important role in black hole physics  such as the Hawking effect \cite{Haw75,ChrFul} and in the dynamics of the early universe,  such as altering the behavior of the universe near the singularity \cite{FHH79},    engendering inflation \cite{Sta80} and addressing the horizon \cite{Anderson} issues. However, for laboratory DCE experiments since the speed of the moving mirror cannot easily be increased to the range where the trace anomaly factor becomes important, this effect is justifiably ignored.  We are not aware of many theoretical papers on the trace anomaly in the context of DCE.  In the few which considered it, e.g., the paper by \cite{2J} on DCE in $S^1$, it was misconstrued as a backreaction effect.  Interested reader can consult a recent paper \cite{XBH} which contains a  trace anomaly calculation and discussions of its backreaction effects in analogue quantum optomechanics (QOM) \cite{QOM}.

Since Casimir effect is familiar to the reader, and trace anomaly is rarely mentioned in DCE, in this paper We shall focus on the most important one of these three quantum processes, namely, particle creation from vacuum fluctuations of the quantum field in both DCE and CPC.  We shall show the primary mechanism they both are based upon, namely, parametric amplification of vacuum fluctuations of a quantum field, and noticeably, their differences, which are mainly in the set-ups, namely, the drive in cosmology is due to the expansion or contraction of the universe, whereas in DCE it comes from an external agent, that which sets the mirror in motion and keeps it in some prescribed trajectory. Notice we use the word ``prescribed" for DCE, but not for CPC. This is because how a universe evolves  follows the solutions of the semiclassical Einstein equation (SCE) \cite{HuVer} with a given quantum matter field source and the scale factor is determined by solving the SCE and the field equation self consistently-- this is referred to as the cosmological backreaction problem. 

\subsection{Seven issues of interest}
In terms of issues we have selected out seven of relevance to discuss.  The first two issues pertain to the \textbf{setup and kinematics}: a) how moving boundaries  change the quantum field enclosed within, as compared to a typical cosmological model.  If we use for DCE the simplest set up of a rectangular box where the sides are allowed to expand or contract, then the analog cosmological model would be the Bianchi type I universe \cite{RyanSheply}, where the scale factors in different directions expand at different rates. We shall explain some subtleties in their  differences, such as, what happens near the boundary?  b) If we use mirrors as examples for DCE, both perfect as with a conductor, or imperfect, as with a dielectric, one can calculate their differences in say, the amount of particles created, e.g., in \cite{Rego}.  Is there a cosmological analogy to an imperfect mirror? Not so straightforwardly. One can consider adding a boundary term in the action for the quantum field, the importance of which for the considerations of how the global structures of spacetime manifest in the quantum field effects was emphasized by Gibbons and Hawking \cite{GibHaw}. This may be an interesting analog gravity issue to explore in the future, but not in our present considerations. However,  for common setups  in DCE experiments it is perhaps more useful to account for the more realistic factors, like  mirrors with finite reflectivity. For this purpose we present the mirror-oscillator-field (MOF) model based on micro-physics processes designed as an economic theoretical framework for quantum opto-mechanics.  It is also useful for expounding dissipation, noise, fluctuations and the fluctuation-dissipation relation which serves as a requisite condition for self-consistency  in backreactions.

The next five issues pertain to \textbf{dynamical processes}, which both DCE and CPC share in an intimate manner, they are: c) nonadiabaticity and parametric amplification; d) enhancement in finite temperature effects is really due to stimulated emission; e) dissipation;   f) backreaction;   g) fluctuation-dissipation relation. 

A word on the style of presentation here. Because the scope of our paper spans two important yet on the surface rather disparate fields which have been active for the past 50 years, we need to be selective on the issues and the coverage. To provide some ordering We have arranged these seven issues from the commonly better known,  mentioned here for completeness,  to the lesser explored yet important issues, some are  currently under investigation.  The extent of our coverage on each issue will go in proportion as well.   Aiming at  a `perspective' for the big pictures, we shall  place more emphasis on the discussions of key ideas.  To economize valuable space, when we invoke formulas in our discussions, we may invoke a few representative papers worthy of some careful study and refer to the formulas therein. Finally, since this paper is in the nature of a `viewpoint', which is almost opposite to a `review' in terms of span and balance, it is almost unavoidable that it draws on the authors' past experiences and refers to their own work of importance. For this we beseech  the reader's patience and forbearance.

\section{Setup: Requisite conditions for the validation of the analogies}\label{S:eoet}
It has been known for many years that the central mechanism of dynamical Casimir or Casimir-Polder \cite{CasPol}  effects in the excitation of the quantum field is the same as in cosmological particle creation, namely, via  parametric amplification, the former by the movement of a mirror or an atom, the latter by the expansion or contraction of the universe. The theoretical developments in particle creation processes in the early universe \cite{Par69,Zel70}, moving mirrors \cite{Moore,DavFul,FulDav} and black holes \cite{Haw75},  have advanced in great strides from the late 60s through the 70s in the Hawking-Unruh effects \cite{Unr76} to the beginning of the 80s in the inflationary cosmology \cite{Guth,Steinhardt,Linde}, pairing with, as their theoretical foundation, the establishment of quantum field theory  in curved spacetime (QFCS) \cite{Fulling,Wald,BirDav,MukWin,ParTom,Ford21} and semiclassical gravity (SCG) \cite{HuVer}.  It would be beneficial to experimentalists and their supporting theorists in the fields of atomic-optical-condensed matter-fluid (AOCF) physics who work on DCE or DCP effects to tap into the useful methodology and  important results from the QFCS and SCG theoretical  enterprises from the 70s to the 90s. It is also helpful to theorists in gravity/cosmology, the present authors certainly included,  to  better understand the AOCF experimental requirements, from their setups to their implementations to the interpretation of their findings, and place them  in the proper contexts of cosmological particle creation to make the best of the many theoretical results obtained before.  This task is what is required in the establishment of a viable analogue gravity program, as witnessed by the highly successful classical fluid \cite{Silke} and BEC experiments \cite{Jack}, to give just two established examples, and many more forthcoming. Please refer to the updated review \cite{AnalogG} for the latest developments.  

It is a happy situation that many prior and ongoing work exist in moving mirror setups linked to the Hawking effect and in accelerating atoms (called `detector'~\cite{Unr76,DeW79} in this community) linked to the Unruh effect \cite{Unr76}. Our emphasis here is on dynamical Casimir (moving mirrors) and  Casimir-Polder (moving atoms near a mirror or medium) effects, not requiring the moving atom or mirror to be uniformly accelerating or to follow any particular trajectory.  The physics of Hawking effect in black holes, the Davies-Fulling effect \cite{DavFul,FulDav} in moving mirrors \cite{Walker,Carlitz,ChuVer,ObaPar,GoodWilczek} and the Unruh effect for accelerating detectors (atoms) are of a different nature from cosmological particle creation,  all three cases in the former group focus on the effects of the event horizons on the  emitted radiation (Hawking effect) or the radiance (an Unruh-DeWitt detector feels hot at the Unruh temperature). Cosmological particle creation originates from the parametric amplification of vacuum fluctuations of the quantum field which does not invoke event horizons.  For de Sitter space, when represented by a spatially-flat ($k=0$) Robertson-Walker universe (the `Poincare patch') expanding at an exponential rate in cosmic time, it provides the inflationary cosmology \cite{Guth} scenario. It falls under our ``cosmology or dynamical spacetime" category, whereas an observer in a de Sitter space in the static coordinatization (see \cite{BirDav} for a display of different coordinatizations of de Sitter space) feeling thermal radiance -- the Gibbons-Hawking effect \cite{GibHaw} -- belongs to the ``black hole" or ``event horizon" category. (Thermal particle creation in both the black hole and  the cosmology categories can be understood as sharing a similar origin, see \cite{HuThermal,KHMR}).  Moving mirror analogue of black hole quantum field processes, Hawking effect and its backreaction has a much broader literature which we will not delve into.   Our discussions here shall focus only on problems and issues in dynamical spacetime of the cosmology category which, by analog gravity correspondence, relate to dynamical Casimir or Casimir-Polder effects.

What we want to do here is to add some solid groundwork to help establishing the linkages between DCE/DCP and CPC to strengthen this  branch of analog gravity program. We try to take the stand of the AOCF in perspective and examine relevant elements in the rich QFCS/SCG theory repertoires which can  reinforce the theoretical foundation of AOCF for the analog gravity purpose. Thus we shall start with the most rudimentary questions, such as, why should we care about `curved' spacetime effects while all of our experiments are carried out in undeniably flat space, and seemingly even more remote, how would topology of space enter.  We address these issues in this section. In the next section we focus on the central mechanism of parametric amplification, resonance and discuss stimulated emission, thermal effects and nonequilibrium issues.  

In the second part of this paper, in Sec.~\ref{S:ebkteo} and \ref{S:bdker},  we move to slightly more complex issues, namely, fluctuations, backreaction and dissipation. We shall bring forth what was  accomplished in SCG, using open quantum system concepts applied to the universe dynamics with quantum fields, to address these theoretical issues in DCE. In Sec.~\ref{S:bdker} we introduce a working model called the M (mirror) O (oscillator) F (field) model for quantum optomechanics, invented a decade ago \cite{GBH,SLH} where a dynamical degree of freedom is assigned to a mirror, so that one could include imperfect mirrors and dielectric medium based on microphysics dynamics. We identify the dynamical variables in DCE and DCP as cogent examples to illustrate  how their interplay  provides a fuller description of the various experimentally accessible effects. Then in the spirit of analogue gravity, we describe a problem in cosmology where a tripartite dynamical variable set-up works well,  namely, the treatment of preheating  in inflationary cosmology. The MOF model is a simple example of a more elaborate and systematic theoretical framework  developed earlier by Behunin and Hu \cite{BehHu10,BehHu11}  known as the {\it graded influence action} formalism based on a nested set of coarse-grained effective actions \cite{cgea}. We give a schematic description without going into the details, but the interested reader can employ this and other related  work using the functional integral formalism \cite{GolKarDCE,HLW,FLM} to explore a host of  complex issues in the nonequilibrium dynamics of the ordered strata of effectively open quantum systems,  such as those manifested in quantum friction \cite{QFric}.  Finally we mention some notable recent papers which further develop the theoretical basis of the MOF model, where their results can be used to describe various aspects of quantum optomechanics of the atom/mirror/field/medium interactions, from motional decoherence \cite{DalNet,ShrHu}, entanglement dynamics \cite{BusPar,RHY} in DCE and in composite particles \cite{Brun} to the detection of quantum features of gravitational waves \cite{qGW,Chen}.

\subsection{Quantum Fields in $S^1$ and $T^3$ Spaces: Boundary conditions and Topology}

DCE has been studied by many authors for more than half a century, see  the excellent review by Dodonov. For expounding the foundational issues let us begin  with the simplest setup of two point mirrors, one located at the origin of a line and the other at distance $l$ from it.  If the mirrors are perfectly reflecting, a condition we assume until we reach  the last section, then a reflecting (Dirichlet) boundary condition on the quantum field at the two ends of this line segment cavity can be imposed. Field theorists  may find it convenient to  replace the line segment by a ring of circumference $l$ by imposing a periodic boundary condition $\phi(x) = \phi (x + l)$ on the quantum field  living on this ring. The range of summation over field modes differ: the Dirichlet condition afforded by a perfectly reflecting mirror  sums from 1 to $\infty$ while the periodic boundary condition sums from $-\infty$ to $+ \infty$. When a line segment is turned into a ring topology changes.  We can think of two aspects in  the {\it physical} similarities and differences between a reflective boundary condition on the two end points and a periodic boundary condition which allows for replacing the line segment by a ring:  

i) Similarities in energy consideration:  The normalization of modes differs by a factor of $\sqrt 2$.  So in terms of actual results these two boundary conditions only differ by a small overall factor.  

ii) Differences in phase consideration: A wave under periodic BC preserves the phase condition, namely, after it traverses the full circle and returns to the origin, it will have the same phase as when it started from the origin.  However, this is not the case for a reflective BC: A wave propagating in the vacuum upon reflection by a mirror suffers  a $\pi$ phase inversion, meaning, a positive crest will turn negative right after reflection\footnote{This depends on the relative strengths of the dielectric medium of the two regions. Here the vacuum of course has a weaker strength than the mirror. In the opposite situation a wave propagating from a denser medium (with larger dielectric constant) to a lighter medium will, upon reflection, keep the same positive phase.}   and this would make an important difference in phase sensitive issues, such as entanglement \cite{Rong}.  

Allowing $l$ to change with time engenders dynamical Casimir effect. Let us introduce a scale factor $a(t)$ which at $t=0$ is set equal to 1. the length of the line segment or the circumference of the circle $l$ has evolved to $  L(t) = a(t)\,l$ at time $t$.   We can then describe the metric of this $1+1$ D spacetime by the line element 
  $ds$  given by ($c=\hbar=1$):
\begin{equation}
    ds^2 \equiv g_{\mu\nu} \dd x^\mu \dd x^\nu = \mathrm{d}t^2 - a(t)^2 \mathrm{d}x^2\,,
    \label{metric}
\end{equation} 
where $g_{\mu\nu}$ is the metric tensor. If $a(t)$ is given a periodic time dependence it becomes a pulsating ring  or sphere.  This motion which preserves spherical symmetry is described as a `breathing' mode in DCE.

To pursue this issue further, we can consider twisted field boundary conditions, e.g., $\phi(x) = - \phi (x + l)$, namely, the field amplitudes have a negative sign  when they make a full circle around. Quantum theories of fields with this type of anti-symmetric BC have been studied by Avis,  DeWitt, Dowker, Isham et al \cite{twisted}.  These issues bearing on the topology of space and fields are addressed in the context of entanglement dynamics on an Einstein cylinder with $S^1$ (space) $\times R^1$ (time) topology in \cite{LCH_S1}.  We will not pursue the effects of field configurations with nontrivial topology here.

Moving on to 3-dimsional space, we meet the popular setup of two parallel conducting plates with time-varying separation \cite{Crocce,Fedotov}. Here, we will be considering a rectangular box whose three faces are made of conducting plates. Recognizing the subtle differences between reflective and periodic boundary conditions on the quantum field we construct the rectangular box by imposing periodic boundary conditions on the field at both ends of each of the three sides. The simplest setup for DCE would be to fix the two transverse directions with equal length and allow the length of the third side (longitudinal direction) to change with time.  A rapid change will engender particle creation from vacuum fluctuations, which is the more ubiquitous and prominent quantum field process in dynamical Casimir effect we choose to focus on here. 


The rectangular box configuration with time dependent sizes mimics that of a symmetric Bianchi Type I universe (with periodic conditions imposed)  with line element
\begin{equation}
    ds^2 \equiv g_{\mu\nu} dx^\mu dx^\nu=  dt^2 - \qty[a_1^2(t)\ dx^2 + a_2^2(t) dy^2 + a_3^2(t)dz^2]\,,
\end{equation}
where $a_i$ are the scale factors in the three directions. We consider the case with $a_2=a_3=1$, $a_1=a(t)$, i.e., with only one side moving, in the $\pm x$ direction.   Imposing periodic boundary conditions at both ends of each of the three sides, i.e., between $x$, $y$, $z = 0$ and $x$, $y$, $z = l$, where $l$ is the coordinate length,  would turn this into a `box universe'  with topology $R\times T^3$ where $R$ refers to the time direction and  $T$ denotes a torus. In our present set up the scale factors in the $y$, $z$ directions are fixed, but the scale factor in the $x$ direction is allowed to change in time.  

On the conditions which validate setting up the DCE-CPC analogy,  we add a note on the issue of spatial homgeneity:  The periodic boundary condition we assume for a system exhibiting DCE preserves spatial homogeneity,  which facilitates direct correspondence with homogeneous cosmology. However, in DCE it is expected that particle creation would be most pronounced in the spatial region closest to the moving mirror as that is the region where the vacuum fluctuations are parametrically amplified  most strongly.  
To draw a closer analogy with homogeneous cosmology cases we need to add this homogeneity assumption for DCE\footnote{The cases in DCE with oscillating mirror are  probably the most commonly described, and second to it, changing the properties of the medium. The case with an infinite medium
in motion without any boundaries and with time independent material constants, which seems closer in features to homogeneous cosmology, has also been discussed, see \cite{Bir2}.}.  Doing so is not too contrived when the external drive is periodic, or at times long compared to the transient after a rapid expansion of the space ends. The field in this space can be considered as homogenized enough to be effectively described by the cosmological analogues used.  The conditions for and the effects of particle creation in both situations (rapidly moving boundary and expanding spacetime) are the same.  It is expected that  after the initial transient has past, the field would approach a stationary state and become spatially homogeneous on average albeit temporally remain in a nonequilibrium steady state. Detailed studies of both DCE and CPC would require nonequilibrium quantum field theory considerations. (See, e.g., \cite{CalHu08}). We shall comment on this in the second part on issues deserving further considerations in future developments.  

\subsection{Coupling between quantum field and spacetime}

We now focus on how a quantum field is coupled to a background spacetime, first a quick review of the familiar QFT in flat Minkowski spacetime, followed by a discussion of a massless conformal scalar field in the Bianchi Type I universe. 

The energy-momentum tensor of a massive scalar field $\phi$ in Minkowski space with metric $\eta_{\mu \nu }$ is
\begin{equation}
    T_{\mu \nu }^{\text{Mink}}=\nabla _{\mu }\phi \nabla _{\nu }\phi -{\frac{1}{2}}\eta_{\mu \nu }\nabla ^{\rho }\phi \nabla _{\rho }\phi -{\frac{1}{2}}\eta _{\mu\nu }m^{2}\phi ^{2}\,.  \label{TmnMink}
\end{equation}
Its energy density is given by the
expectation value of the 00 component of $T_{\mu \nu }$ with respect to the Minkowski vacuum, i.e.,
\begin{equation}
    \rho _{0}^{\text{Mink}}\equiv \langle 0\vert T_{00}\vert 0\rangle =\int\! \frac{d^{3}\mathbf{k}}{2(2\pi )^{3}}\;\bigl(\lvert \dot{f}_{\mathbf {k}}\rvert ^{2}+\omega _{\mathbf {k}}^{2}\,\lvert f_{\mathbf {k}}\rvert ^{2}\bigr)=\int\! \frac{d^{3}\mathbf{k}}{(2\pi )^{3}}\;\bigl(2s_{\mathbf {k}}+1\bigr)\frac{\hbar \omega _{\mathbf {k}}}{2}\,.  \label{rho0Mink}
\end{equation}
where $f_{\mathbf {k}}$ are the amplitude functions of the $\mathbf{k}^{\text{th}}$ normal mode of the field. They obey an equation of motion like that of parametric oscillators, meaning, whose frequencies are time-dependent.
The Hamiltonian for the dynamics of a collection of  parametric oscillators is given by
\begin{equation}
    H^{\text{Mink}}(t)=\frac{1}{2}\sum_{\mathbf{k}}\bigl(\pi _{\mathbf {k}}^{2}+\omega _{\mathbf {k}}^{2}q_{\mathbf {k}}^{2}\bigr)=\sum_{\mathbf{k}}\Bigl(N_{\mathbf {k}}+\frac{1}{2}\Bigr)\,\hbar \omega _{\mathbf {k}}\,,
\end{equation}
Comparing this with Eq. (\ref{rho0Mink}) one can identify $\lvert f_{\mathbf
{k}}\rvert^{2}$ and $\lvert \dot{f}_{\mathbf {k}}\rvert ^{2}$ with the canonical coordinates $q_{%
\mathbf {k}}^{2}$ and momenta $\pi _{\mathbf {k}}^{2}$, and the eigenvalue of $H_{0}$
is the energy $E_{\mathbf {k}}=(N_{\mathbf {k}}+1/2)\,\hbar \omega _{\mathbf {k}%
}$. 
The analogy of particle creation with parametric amplification is formally made clear
\begin{equation}
    N_{\mathbf {k}}(t)+\frac{1}{2}=\frac{1}{2\hbar \omega _{\mathbf {k}}}\bigl(\pi _{\mathbf {k}}^{2}+\omega_{\mathbf {k}}^{2}q_{\mathbf {k}}^{2}\bigr)\propto s_{\mathbf {k}}\,,  \label{Nk}
\end{equation}
Please refer to Eq. (\ref{sk}) which gives the number density and Eq. (\ref{rho0Mink}) says that the energy density of vacuum particle creation comes from the amplification of vacuum fluctuations $\hbar \omega _{%
\mathbf {k}}/2$ by the factor $\mathcal{A}_{\mathbf {k}}=2s_{\mathbf {k}}+1$. Now it is
easy to recognize the Minkowski result:
Since there is no particle creation $s_{\mathbf {k}} = 0$   there is no amplification $%
\mathcal{A}_{\mathbf {k}}=1$,  
In the next section we shall see how particle creation arises when the background spacetime becomes dynamical. 

We now consider a massless conformal quantum field in the rectangular box with one face moving, in analogue to a symmetric Bianchi Type I universe with periodic boundary conditions imposed on three sides, as described earlier. (Due to the imposition of boundary conditions the global topology of our setup is not the same as the Bianchi Type I Universe, it becomes $T^3$, a three-torus.) We wish to explain why the curvature scalar $R$ appears in the field equation.  This is a simplified model of electromagnetic field, which is conformal (photons are massless and the Maxwell equation in four spacetime dimension is conformally-invariant) -- simplified here in our treating just one of the propagating components of a vector field, ignoring the polarizations, which does not alter the parametric amplification mechanism in any significant way\footnote{This may affect the kind of DCE arising from an atomic polarizability-dependent parametric interaction with the field. See \cite{LoLaw}}.  The answer to this query is, indeed the space is flat, so the intrinsic curvature is zero, but because one or more faces are moving, the extrinsic curvature term is nonzero. We shall explain some more after we show the expression for the wave equations governing the normal modes of the field.

Consider a massless conformal scalar field $\phi$  in this box `universe'  with action
\begin{equation}
    S_f = \frac{1}{2}\int\! dx^4\;\sqrt{\lvert g\rvert} \,\qty(g^{\mu\nu}\,\partial_{\mu}\phi\,\partial_{\nu}\phi -\frac{R}{6}\,\phi^2)\,,
\end{equation}
obeying the Klein-Gordon equation,
\begin{align}
    \Box\phi +\frac{R}{6}\phi &= 0\,, &&\Rightarrow &\partial_t^2\phi + \frac{\dot{a}}{a}\,\dot{\phi} - \sum_i\frac{\partial_i^2\phi}{a_i^2}+\frac{R}{6}\,\phi&=0\,, 
\end{align}
where $R$ is the scalar curvature of spacetime the field lives in. Here comes perhaps the first legitimate question: For the box universe, the space is flat so where could the curvature enter?

Examining the form of this curvature scalar for a symmetric Bianchi Type I universe, we can see it is proportional to $\ddot{a} (t)$, the acceleration in time of the scale factor. This is the extrinsic curvature we referred to above, which is nonzero for a  time-dependent space, as in dynamical Casimir and Casimir-Polder effects. There is no such contribution if the space is static and flat (zero intrinsic curvature), as in Minkowski space shown before\footnote{When a researcher in general relativity mentions quantum fields in a “curved” spacetime, that context includes also dynamical flat spaces. The scalar curvature has two contributions: even though the intrinsic curvature of flat space is zero, the extrinsic curvature measuring the time rate of change of the scale factor (here the moving boundary) is nonzero. This is the same situation as a spatially-flat Robertson-Walker (RW) universe being considered as curved spacetime: spatially-flat has zero intrinsic curvature, yet expanding space makes it “curved”,  whereas a closed or open RW universe has both intrinsic and extrinsic curvatures. }   

Proceeding forward, the energy-momentum tensor of this scalar field is given by
\begin{align}
    T_{\mu\nu} \equiv  { \frac{2}{\sqrt{\lvert g\rvert }}\frac{\delta S}{\delta g^{\mu\nu}}} &= (\partial_\mu\phi)(\partial_\nu\phi)-\frac{1}{2}g_{\mu\nu}g^{\lambda\sigma}(\partial_\lambda\phi)(\partial_\sigma\phi) \nonumber\\
    &\qquad\qquad-\frac{1}{6}\left[\nabla_\mu\partial_\nu(\phi^2)-g_{\mu\nu}g^{\lambda\sigma}\nabla_\lambda\partial_\sigma(\phi^2)+\phi^2\,G_{\mu\nu}\right]\,,  \label{Tmn}
\end{align} 
where $G_{\mu\nu}\equiv R_{\mu\nu}-\frac{1}{2}g_{\mu\nu}R$ is the Einstein tensor. This expression for the stress-energy tensor,  introduced by Callen, Coleman and Jackiw \cite{CCJ} (who  called it the `new and improved'),  is the correct one to use for conformal fields. Notice that it contains the Einstein tensor. Even though in the common representation  of electromagnetic theories the background spacetime structure is rarely mentioned,  it actually plays an important dynamical role in the determination of the stress energy tensor, even in flat spacetimes with nonzero extrinsic curvature. 

For a quantum field, the stress-energy tensor is upgraded to be an operator. Taking the vacuum expectation value of the $00$ component gives the  energy density, which is the quantity of interest for particle creation studies:   \begin{equation}
    T_{00}=\frac{1}{2}(\dot{\phi})^2 + \frac{1}{3}\frac{\dot{a}}{a}\,\phi\dot{\phi}+\frac{1}{6}\sum_i \left(\frac{\partial_i\phi}{a_i}\right)^2-\frac{1}{3}\sum_i \left(\frac{\phi\partial_i^2\phi}{a_i^2}\right)-\frac{1}{6}G_{00}\,\phi^2\,,  \label{T00}
\end{equation}
where an overdot indicates differentiation with respect to $t$.  We shall continue this exposition in the next section when we discuss particle creation as a nonadiabatic process.  

\subsection{Factors in DCE and cases in cosmology with finer correspondences}

\subsubsection{Cosmology of and DCE in 3D space}   
Before closing this section we mention cases where the correspondences between DCE in conventional setups (such as parallel plates or conducting spheres) and CPC are not so well placed. We then suggest alternative analogue setups which we see as having better promises for continual developments.   

For example, the closed FLRW universe is an important member of the Standard Model of cosmology, but there is no such correspondence in the set up of DCE in 3-dim space. In DCE, a popular model is that of a pulsating spherical shell, but (in the limit the shell is infinitely thin) that is really a $S^2$ geometry with a time-dependent scale factor. The mathematical definition of a sphere refers to its surface, not the interior, known as a  `ball' (or a disk in 2D, as different from its perimeter, the ring, in 1D).   DCE is interested in the effects of quantum fields in the space contained within the moving shell, namely the three dimensional ball $B^3$ ($B$ for ball). The closed FLRW universe refers to the whole spatial extent of a three sphere $S^3$ with time-dependent radius, not the space within a spherical shell $S^2$, which is a three dimensional ball $B^3$.  

Having said this, one may be able, but not necessarily convenient, to set up   some analogy between a pulsating shell in DCE with gravity, not with cosmology per se, but with  black holes. The membrane paradigm \cite{ThorneMembrane} may be a good entry point, because the intricacies of a black hole related to its event horizon are mapped into physical properties of a membrane. For massive stars, observing DCE originating from a vibrating boundary in the exterior of a Schwarzschild metric was mentioned in \cite{Ive}.

If one does not insist on moving plates or pulsating shells, Bose–Einstein condensate (BEC) is a more malleable and richer medium. Indeed, serious attempts have been made to set up correspondences between what goes on in the early universe and specific dynamics of BEC, leading to what's called `laboratory cosmology', e.g., \cite{CalHuLabCos,TedIanGret}. Many fine experiments have mapped out the controllable parameters, e.g., the time-varying quantity is the scattering length via a Feshbach resonance.  Because BEC is formed from atoms, theoretical analyses on its dynamical responses can reveal hard to access information about its underlying substructure, e.g., the quantity in BEC corresponding to the Planck length in cosmology is the healing length. There are many DCE experimental proposals using changing or `modulated' BEC, see, e.g., \cite{Iacop,Demler,MotDal}. A more complete list of references on DCE experiments with BEC can be found in    \cite{Dodonov} Sec. 4.2. Please also consult updates of reviews on analogue gravity \cite{AnalogG} for the latest developments .

\subsubsection{Imperfect mirror}
The reflectance of a mirror certainly is a very important factor to consider in  the design of  quantum optics experiments to meet high precision demands.  For a perfect conductor one can just impose a boundary condition on the field at the position of the mirror. What are the analogues in cosmology? In Euclidean quantum gravity Gibbons and Hawking asserted the importance of boundary terms in the functional integration over equivalent classes of conformal spacetimes. In particle physics, higher-dimensional, especially braneworld cosmology, what happens in the boundary occupies the main stage. The essence of AdS (anti-de Sitter space)-CFT (conformal  field theory) physics hinges on the existence of a duality relation between gravity in the bulk and  gauge theories on the boundary. One can in principle think of atomic-optical-condensed matter analog experiments to check on the theoretical predictions by these theories.  We will not pursue these fancier lines of thought here, but just mention a down to earth model known as the mirror-oscillator-field (MOF) model for the description of imperfect mirrors. There, the internal degrees of freedom of an atom or a mirror is made a dynamical variable which can interact with the field. How the dynamics of this variable determines the reflectivity of a mirror has been calculated in these original papers \cite{GBH,SLH,LinMOF}. We shall return to the structure and functions of the MOF model in the last section.

\section{Mechanism, Dynamics and Processes}
The above discussions are about the setups, from conditions on the quantum fields (boundary conditions) to the characters of spacetimes (curvature and topology) and the nature of their couplings. 
We now turn to dynamics, first focusing on the key mechanism of particle creation, namely, nonadiabatic processes and parametric amplification, then generalize from vacuum to finite particle initial states in stimulated particle production, and to finite temperature effects, ending with consideration of the fully nonequilibrium dynamics.

\subsection{Particle creation is a nonadiabatic process, negligible under adiabatic approximations} 
The equation of motion for the amplitude function  $f_{\mathbf{k}}$ of the $\mathbf{k}^{\text{th}}$ normal mode of a quantum field in a dynamical spacetime written in a generic form is that of a harmonic oscillator with time-dependent frequency $\omega_{\mathbf{k}}(t)$ :
\begin{equation}
	\ddot{f}_{\mathbf{k}} + \omega_{\mathbf{k}}^2(t) f_{\mathbf{k}} = 0\,.
\label{Eq:hk}
\end{equation}
(An overdot denotes differentiation with respect to time $t$).
For the treatment of particle creation in a symmetric Bianchi Type I universe with periodic boundary conditions, as the analog of DCE in a rectangular box with one moving face, it is more transparent to use the conformal time $\eta$ and the conformally-related field $\chi(\eta)$  
\begin{align}
    \chi &= a^{1/3}\phi\,, &\dd\eta &=  a^{-1/3}{\dd t}\,,
\end{align}
resulting in an equation for the mode function $\chi_{\vec{k}}(\eta)$
\begin{equation}
    \chi_{{\mathbf{k}}}''+(\Omega_{{\mathbf{k}}}^2+Q)\,\chi_{{\mathbf{k}}}=0\,.
    \label{xeq}
\end{equation}
(A prime refers to taking the derivative with respect to the conformal time $\eta$) with a time-dependent frequency 
\begin{equation}
    \Omega_{{\mathbf{k}}}^2=\qty(a^{1/3}\omega_{{\mathbf{k}}})^2=a^{2/3}\sum_i\qty(\frac{k_i^2}{a_i^2}+m^2)\,.
\end{equation}
Here, $Q[a'_i(\eta)]$ is defined by 
\begin{equation}
    Q = \frac{1}{18}\sum_{i<j}\qty(\frac{a_i'}{a_i}-\frac{a_j'}{a_j})^2\,.
\end{equation}
It measures the degree of anisotropy in the expansion of the spacetime and is  notably independent of ${\mathbf{k}}$.


The region in $\mathbf{k}$-space which yields the most particle production contains  normal modes which are subjected to strong nonadiabatic parametric amplification. As a qualitative measure for each $\bf k$ mode one can use the nonadiabaticity parameter introduced in  \cite{Hu74} (Eq.73-74):   $\bar \omega \equiv \dot\omega_{\mathbf{k}}/ \omega^2_{\mathbf{k}}$ in cosmic time (Minkowski time here) or $\bar \Omega \equiv \Omega'_{\mathbf{k}}/\Omega^2_{\mathbf{k}}$ in conformal time.
Those fulfilling the condition $\bar \omega \gg 1$ in cosmic time or $\bar \Omega \gg 1$ in conformal time are the modes which contribute most to particle creation.  Integrating over these   non-adiabatic modes in the energy density expression  yields the dominant contribution to particle creation.

This true statement is probably well-known amongst the practitioners.  But in the literature there were cases where processes based on the same mechanism as DCE or CPC were treated by adiabatic approximations, which led to qualitatively opposite results and wrong estimates. An example is the proposal of sonoluminescence (SN) by Schwinger \cite{SchSN} as quantum vacuum radiation, followed by the work of Eberlein \cite{Eberlein} and others\footnote{Comments doubting the viability of DCE being the primary cause of SN followed soon after, e.g. \cite{Lambrecht} claiming that  fundamental physical limitations on bubble surface velocity limit the predicted number of photons per flash to be much smaller than unity; and  \cite{UnnMuk} pointing out the importance of the details of the trapped gas, that there is no SN if there us no trace of inert gases in the trapped air, and the dependence of the intensity on the ambient temperature.}. The prevailing explanation of this interesting effect today seems to lean towards a plasma physics phenomenon. The controversy of Schwinger's original proposal, see, e.g., \cite{MiltonSN},  is not what we want to delve into here.  There were efforts to distinguish the mechanism responsible for SN, e.g.,  Sciama and co-workers \cite{SciamaSN} pointed out the difference between what they call `real' thermal light and the `pseudo'-thermal squeezed-state photons typical of models based on the dynamic Casimir effect. They show that two-photon correlations provide a powerful experimental test for various classes of sonoluminescence models.  Granting for the sake or argument that SN is a kind of quantum radiation resulting from the parametric amplification of vacuum fluctuations, with the same mechanism as in DCE,  the point we are making here is that,   using ``standard perturbation theory with an adiabatic approximation" is  not the way to go: Particle creation is a nonadiabatic process. Also, it is a rather generic feature that for high frequency ${\mathbf{k}}$ modes the number of particles produced falls off exponentially with ${\mathbf{k}}$ (see, e.g., Eq. (3.125) of \cite{BirDav}) which explains why ``the emitted light resembles a black body spectrum." A recent paper \cite{KarMai} calls for a reexamination using exact solutions, which we support -- the methodology has been laid out much earlier in the study of cosmological particle creation.

A common reason for such missteps probably comes from the habitual practice of invoking the WKB approximation as the first line of attack to solving a differential equation with no exact analytic solutions.  Let us try this out, i.e., invoke a WKB approximation to the equation of motion for the amplitude functions of the normal modes and see what we come up with.  We first derive a formal expression for the number of particles  $s_{\mathbf {k}}\equiv \lvert \beta _{\mathbf {k}}\rvert^{2}$ (where $\beta_{\mathbf{k}}$ is the Bogoliubov coefficient measuring the portion of negative frequency component evolved from an initial state containing only positive frequency component) produced in terms of the amplitudes of the wave function $\lvert f_{%
\mathbf {k}}\rvert$ and $\lvert \dot{f}_{\mathbf {k}}\rvert$. We then carry out a WKB expansion to arrive at expressions for $s_{\mathbf{k}}$ in successive  adiabatic orders. The following is excerpted from \cite{CalHu08} which is based  on \cite{ZelSta,Hu74}.

We seek a solution in the form:
\begin{align}
    f_{\mathbf {k}}(t)&=\sqrt{\frac{\hbar }{2\omega _{\mathbf {k}}}}\left( \alpha _{\mathbf {k}}e_{\mathbf {k}}^{-}+\beta _{\mathbf {k}}e_{\mathbf {k}}^{+}\right)\,, &e_{\mathbf {k}}^{\pm }\equiv \exp \left(\pm i\int\!dt\; \omega _{\mathbf {k}}\right)\, .
\label{chialbe}
\end{align}
The two functions $\alpha _{\mathbf {k}}$, $\beta _{\mathbf {k}}$ are
the positive and negative frequency components of a formal solution
$f_{\mathbf {k}}$, but without a well-defined vacuum they do not
convey the meaning of particles and antiparticles, as we forewarned
with regard to the Bogoliubov coefficients. Since the single equation
Eq.~(\ref{chialbe}) does not determine the coefficients $\alpha _{\mathbf
{k}}$ and $\beta _{\mathbf {k}}$ uniquely, we need another condition. We shall choose it such that  the Wronskian condition 
$\lvert \alpha _{\mathbf{k}}\rvert^2 - \rvert\beta _{\mathbf {k}}\rvert^2 = 1$, 
which guarantees the unitarity of the system dynamics under Bogoliubov transformations, is satisfied. The auxiliary condition imposed on $\dot{f}_{\mathbf {k%
}}$ is
\begin{equation}
    \dot{f}_{\mathbf {k}}(t)=-i\sqrt{\frac{\hbar \omega _{\mathbf {k}}}{2}}\left(\alpha _{\mathbf {k}}e_{\mathbf {k}}^{-}-\beta _{\mathbf {k}}e_{\mathbf {k}}^{+}\right)\, .
\label{fdotab}
\end{equation}
Inverting these two equations we can express the complex function $\beta _{%
\mathbf {k}}$ in terms of $\lvert f_{\mathbf {k}}\rvert^{2}$, $\lvert \dot{f}_{\mathbf {k}}\rvert^{2}$ as follows:
\begin{align}
    \alpha _{\mathbf {k}}&=\sqrt{\frac{\omega _{\mathbf {k}}}{2\hbar }}\left( f_{\mathbf {k}}+\frac{i}{\omega _{\mathbf {k}}}\dot{f}_{\mathbf {k}}\right) e_{\mathbf {k}}^{+}\,, 
    &\beta _{\mathbf {k}}&=\sqrt{\frac{\omega _{\mathbf {k}}}{2\hbar }}\left( f_{\mathbf {k}}-\frac{i}{\omega _{\mathbf {k}}}\dot{f}_{\mathbf {k}}\right) e_{\mathbf {k}}^{-}\,.
\label{albeffd}
\end{align}
Making use of the Wronskian condition we obtain
\begin{equation}
    s_{\mathbf {k}}\equiv \lvert \beta _{\mathbf {k}}\rvert ^{2}=\frac{1}{2\hbar \omega _{\mathbf {k}}}\left( \lvert \dot{f}_{\mathbf {k}}\rvert ^{2}+\omega _{\mathbf {k}}^{2}\lvert f_{\mathbf {k}}\rvert ^{2}\right) -\frac{1}{2}\,.  \label{sk}
\end{equation}
It is tempting to regard $s_{\mathbf{k}} = \lvert \beta_{\mathbf{k}}\rvert^2$ as the amount of particle production. However, we need to be careful that the vacuum state is well-defined to make sense of particles. We shall leave out the construction of the nth order adiabatic vacuum as it is well treated in textbooks where one can trace back to the original papers.  

Now we perform a WKB expansion on   $s_{\mathbf{k}}$ and obtain the second and forth order adiabatic expressions for  $s_{{\mathbf{k}}}$:
\begin{align} 
    s_{{\mathbf{k}}(2)} &=\frac{1}{16}\bar{\omega}^{2} \,,\\
    s_{{\mathbf{k}}(4)} &=\frac{1}{16}\left( -\frac{1}{2} \frac{\bar{\omega}\bar{\omega}''}{\omega ^{2}}+{\frac{1}{4}}\frac{\bar{\omega}'^{2}}{\omega ^{2}}+\frac{1}{2} \frac{\bar{\omega}'\bar{\omega}^{2}}{\omega }+{\frac{3}{16}}\bar{\omega}^{4}
 \right)\,.
\end{align}
In a suitably defined adiabatic vacuum, the above expressions provide the number of particles created from vacuum fluctuations {\it under the adiabatic approximation}, which corresponds to letting $\bar\omega \ll 1$. We can see that particle creation obtained under the adiabatic approximation is very small\footnote{The adiabatic expansion for particle production in the high frequency range at the zeroth, second and fourth adiabatic order above matches the quartic, quadratic and logarithmic divergences in the vacuum energy density respectively. Substituting these expressions for $%
s_{{\mathbf{k}}(\text{div})}=1+s_{{\mathbf{k}}(2)}+s_{{\mathbf{k}}(4)}$ for each $\mathbf {k}$ mode into the expression for the vacuum energy density 
one can identify the divergent contributions. This is the basis for adiabatic regularization \cite{AdiabReg}. Note it is letting ${\mathbf{k}}$ go to infinity which causes the UV divergence, not because the particle number gets very high.  The identification of successive divergent terms is facilitated by an {\it adiabatic} expansion of the particle production expression.}. This is the reason why one should not invoke any adiabatic approximation for the calculation of particle production because it is intrinsically a nonadiabatic quantum process.


\subsection{Enhancement at finite temperature comes from stimulated emission, true only for bosons} 
What we have discussed so far concerns particle creation from the vacuum. We learned that it is due to the parametric amplification of vacuum fluctuations of the quantum field at zero temperature. What happens at finite temperature? Plunien et al \cite{Ralf} studied the thermal effects on the creation of particles under the influence of time-dependent boundary conditions.   For a resonantly vibrating mirror cavity the thermal effect on the number of created photons was obtained nonperturbatively. These authors concluded that finite temperatures can enhance the pure vacuum effect by several orders of magnitude. 

Finite temperature effects certainly are important because most experiments are not carried out at zero temperature. We wish to add some comments to a) identify the root cause of this effect  and b) reveal the crucial condition their claim depends on.  

a) {\it A first discriminant: enhancement is possible only for bosons, not  for fermions}. Fermions may see an attenuation~\cite{Pa71,BA88,SS21,ML21,JW22}.

b) {\it The root cause and more general underlying mechanism is stimulated emission,} namely, the parametric amplification of a state already with $N$ particles present. How these particles, as long as they are bosons, are initially distributed is secondary. The greater amount of particles produced found in Plunien et al comes about not because they obey a thermal distribution, thus regarded as  a finite temperature effect, but because they are parametrically amplified in a stimulated production process.

Let us first discuss stimulated production, then thermal effects. 
If at the initial time there are already   $N_{\mathbf {k}}(t_{0})$ numbers of particles present then the same parametric amplification mechanism which, when applied to vacuum fluctuations gives rise to spontaneous particle production as we saw before, now gives rise to stimulated production. It yields an energy density $\rho _{n}$ with respect to the $n$-particle state at $t_{0}$ given by
\begin{align}
    \rho _{n}^{\text{Mink}}=\langle n\vert T_{00}\vert n\rangle &=\int\! \frac{d^{3}\mathbf{k}}{(2\pi )^{3}}\;\bigl(\lvert\dot{f}_{\mathbf {k}}\rvert ^{2}+\omega _{\mathbf {k}}^{2}\lvert f_{\mathbf {k}}\rvert ^{2}\bigr)\,\langle \hat a_{\mathbf {k}}^{\dagger}\hat a_{\mathbf {k}}\rangle \,,\notag \\
    &=\int\!\frac{d^{3}\mathbf{k}}{(2\pi )^{3}}\;\bigl(2s_{\mathbf {k}}+1\bigr)\,\hbar \omega _{\mathbf {k}}\,\langle \hat N_{\mathbf {k}}(t_{0})\rangle\, .  \label{rhonMink}
\end{align}
Comparing this expression with Eq. (\ref{rho0Mink}) we see that the  
$\frac{1}{2}\hbar \omega_{\mathbf {k}}$ term is replaced by  
$N_{\mathbf {k}}(t_{0})\,\hbar \omega_{\mathbf {k}}$, the number of particles originally present in the $\mathbf {k}$th mode. It is multiplied by an amplification factor. As mentioned earlier, in a Minkowski spacetime, there is no particle creation, $s_{\bf k} =0$, and the amplification factor is equal to one. An expanding spacetime would produce particles from the vacuum fluctuations of the field and from particles originally present. 
Combining Eq. (\ref{rho0Mink}) and Eq. (\ref{rhonMink}), for a density matrix diagonal in the number state, the total energy density of particles created from the vacuum and from those already present in the $n$-particle state is given by
\begin{equation}
    \rho ^{\text{Mink}}=\rho _{0}^{\text{Mink}}+\rho _{n}^{\text{Mink}}= \int \!\frac{d^{3}\mathbf{k}}{(2\pi )^{3}}\;\bigl(2s_{\mathbf {k}}+1\bigr)\,\hbar \omega _{\mathbf {k}}\left[\frac{1}{2}+\langle N_{\mathbf {k}}(t_{0})\rangle\right]\,.
\end{equation}

For bosons obeying a thermal distribution at temperature $T=\beta ^{-1}$, magnification of the $n$-particle thermal state gives the finite-temperature contribution to particle creation, with energy density
\begin{equation}
    \rho _{T}^{\text{Mink}}=\int\! \frac{d^{3}\mathbf{k}}{(2\pi )^{3}}\;\bigl(2s_{\mathbf {k}}+1\bigr)\,\frac{\hbar \omega _{\mathbf {k}}}{e^{\beta \hbar \omega _{\mathbf {k}}}-1}\,.
\label{rhoTMink}
\end{equation}
We can see that the large enhancement in the amount of particles produced as reported in Plunien et al is more due to stimulated amplification of existing particles than as a finite temperature effect per se.  The former refers to parametric amplification of a state already with $N$ particles present, the latter refers to the particular distribution they find themselves in, such as the B-E distribution for Bose particles.  Finding a greater number of particles produced as compared to the vacuum initial state applies to any state with a finite number of bosons present initially, it does not matter whether the n particles were in equilibrium, obeying a thermal distribution such as the Bose-Einstein distribution. Hence it is not a thermal effect per se.  

As a matter of fact, the thermal equilibrium condition these particles find themselves in originally may not hold in a dynamical setting. This is because at every moment there are newly produced particles which will be parametrically amplified. The system at any moment is in a nonequilibrium condition. One needs to examine scattering processes amongst the produced particles to assess how fast the system can settle into equilibrium, and that depends on how fast the system is being driven compared to the rate of these processes, and many other factors. Equilibrium condition cannot be assumed to exist or be maintained without such scrutiny.

To conclude our discussion on this topic of thermal particle production in DCE and cosmology, what was found in Plunien et al, namely, the enhancement of particles produced at finite temperature compared to that from the vacuum, is not a thermal effect per se,  the root cause for enhancement is stimulated emission of bosons, it is not true for fermions. 

Now, let's see how this formulation of finite temperature field theory appears in cosmology. For a conformally-invariant field like photon in a conformally-flat spacetime like the Friedmann-Lamaitre-Robertson-Walker (FLRW) universe, we would obtain the same expressions for the conformally-related field $\chi$ in conformal time $\eta$ (except for a scaling factor of ${1}/{a^{4}}$)  
\begin{equation}
    \rho _{T}^{\text{conf}}=\frac{1}{a^{4}}\int \!\frac{d^{3}\mathbf{k}}{(2\pi )^{3}}\;\bigl(2s_{\mathbf {k}}+1\bigr)\,\frac{\hbar \omega _{\mathbf {k}}}{e^{\beta \hbar \omega _{\mathbf {k}}}-1}\,.
\label{rhoTconf}
\end{equation}
If $s_{\mathbf {k}}=0$, meaning, if there were no new particles created during the evolution -- and this is related to the condition for maintaining thermal equilibrium,  the Stefan-Boltzmann relation holds for a massless conformal field in FLRW universe
\begin{equation}
    \rho _{T}^{\text{conf}}=\frac{\pi ^{2}}{30\hbar ^{3}}\,T^{4}\,.
\end{equation}
Thus $Ta$ is a constant throughout the evolution of the radiation-dominated
FLRW universe. $N_{\gamma }\sim \left( Ta\right) ^{3}$ is proportional to the
number of relativistic particles present which provides the entropy content of the universe \cite{Hu82}. 

Further discussions of finite-temperature particle creation, nonequilibrium conditions \cite{Hu83} and the related entropy generation problem can be found in \cite{Landi,HHEntropy}.



\section{Backreaction: Vacuum viscosity, Quantum Lenz law and Fluctuation-Dissipation relation}\label{S:ebkteo}

\subsection{Dissipation due to the backreaction of particles created by the parametric amplification of quantum fluctuations has memory}  
This long title summarizes the logical sequence of events and effects (particles are created) and processes (via parametric amplification), beginning with the quantum fluctuations of the field. In this section we consider the eventful duo-elements: backreaction (process) and dissipation (effect).

We learned from the last section that a rapid expansion of the universe can parametrically amplify the vacuum fluctuations of a quantum field into particle pairs. Here we study their effects on the dynamics of the early universe. This is known as the backreaction problems. The theoretical basis is semiclassical gravity \cite{HuVer}, where  both the semiclassical Einstein (SCE) equation  and the quantum field equation of motion need be solved in a self-consistent manner.  The SCE equation has as source, in addition to the classical stress energy tensor which we, focusing  on purely quantum effects, will ignore,  also the expectation values of the  stress energy tensor operator of a quantum field $\phi$, in the form of the Einstein tensor  $G_{\mu\nu}[g_{\alpha\beta}] = 8 \pi \mathrm{G}\, \langle \hat T_{\mu\nu} (\phi)\rangle$ where $\mathrm{G}$ is Newton's constant.

In most situations the backreaction of quantum processes like particle creation engenders dissipation in the system's dynamics (an interesting exception is the quantum effects of trace anomaly, see \cite{XBH}).  To address this issue it is useful to adopt some concepts and techniques in open quantum systems. For cosmology, the open system is the evolving universe, for DCE, the moving mirror, while in both cases, their environment is the quantum field. Since backreaction effects in DCE \cite{BkrnDCE} and related phenomena in analog black hole experiments \cite{Jack} are gradually surfacing to the fore it may prove useful to describe what had been done in early cosmology, so we can benefit from the wisdom gained there to better formulate accessible problems of interest for the exploration of similar issues in analogue gravity experiments. 

We shall present results from one of the best studied cases of cosmological backreaction problem, that of particle creation of massless conformal scalar fields in a Bianchi Type I universe, for two reasons:  1) The effect is strong and can lead to a rapid isotropization of the early universe near the Planck time. 2) Many methods have been developed and applied to this problems, e.g, it has been treated via a) both the canonical quantization \cite{HuPar78} and the path-integral methods \cite{HarHu79}, b) both the in-out and in-in or closed time path, Schwinger-Keldysh \cite{SchKel} methods \cite{CalHu87} and c) the Feynman-Vernon formalism \cite{FeyVer} open quantum system methods \cite{HuSin95}, which is particularly useful for understanding of the properties of quantum noise. With part c) we begin to enter into the realm of semiclassical stochastic gravity \cite{HuVer}, where the two point correlations of the stress-energy tensor becomes the centerpiece.  If one's interest is in nonlocal dissipative  and non-Markovian (memory) effects one can stay within the realm of semiclassical gravity and focus on Eq. (3.18) or (3.24) in \cite{CalHu87} for a closer scrutiny. To see how colored (some communities call this nonMarkovian) noises associated with quantum fluctuations of the field enter, focus on Eq. (5.13) of \cite{HuSin95}, where a fluctuation-dissipation relation for particle creation from the parametric amplification of the {\it quantum fluctuations} of the matter field and its {\it dissipative backreaction} on spacetime dynamics is derived for semiclassical cosmology.   

With this background, we can resume our studies of DCE in a symmetric rectangular box with one face moving, starting with the formula for the stress-energy tensor of a massless conformal scalar quantum field Eq. (\ref{Tmn}). Of special interest  is the energy density Eq. (\ref{T00}). These expressions have UV divergences and require viable schemes of regularization or renormalization. This quest occupied the attention of most researchers in QFCS in the 5 years from 1974-78, covering half a dozen methods, four of them remain in  continual usage. They are, the dimensional, zeta-function, point-separation and adiabatic regularization, the last one being the most adept for cosmological particle creation considerations. Serious studies of backrection began in 1978, continued on in the next decade, and finished off with the establishment of semiclassical gravity.

We shall not repeat describing the procedures of regularization nor the assortment of different parts of the energy density responsible for different physical processes involved. Please read Sec. 3 of \cite{XBH} for the details. Noteworthy is a general principle evoked from these results, first discovered for the antecedent cosmological backreaction problems \cite{LenzLaw}. Called the `{\bf Quantum Lenz Law}' by Hu, it says, similar to the classical law of magnetic induction,  that the backreaction will act in such a way as to resist further changes. The findings in \cite{XBH} validate  its manifestation for DCE, at least for the case of a moving rectangular box studied there in detail, and we believe it applies broadly as a law. 

\subsection{Quantum Lenz law, vacuum viscosity, energy balance, self-consistency condition and fluctuation-dissipation relation}

The quantum Lenz law encapsulates several interesting features of backreaction effects of quantum processes, here exemplified by the effects of particle creation from vacuum fluctuations in DCE or cosmology. We will mention three. First,  the backreaction force exerted by the created particles, which results from the quantum field being parametrically amplified, is not just a dissipative force, it is also reactive, like the combined effects of resistance and inductance.  Second, {\it nonMarkovian dynamics, memory}, where the response has a time lag or phase difference.  Here, the nonlocal dissipation kernel on the LHS of the Langevin equation is a clear sign of the existence of memory:    what happens at one particular moment in time depends on the history of its past -- how far back is a measure of the degree of nonMarkovianity. 

{\it Memory effects} in an open system is unavoidable when backreaction is included in the consideration. The physical reason is easy to see: the system, be it the moving mirror or the evolutionary universe, and its environment, here the quantum field, each follows its own dynamics described by an ordinary differential equation with disparate characteristic time scales. When one coarse-grains the information of the environment and absorbs its overall effect into the system, the open system will be described by an integral-differential equation with a nonlocal kernel -- nonlocal because it contains the time scales of both  subsystems to begin with.

Third, {\it energy balance} The energy extracted from the field derived from taking the energy density of particles produced and integrating over all modes covering all space is supplied in the case of cosmology by the expansion of the universe, according to Einstein's equations. The semiclassical Einstein equation which governs   the dynamics of both the spacetime and the quantum field is self-regulating because its driving source, the expectation value of the stress-energy tensor of the scalar field, depends on how the field changes, which in turn obeys a field equation on a background spacetime whose dynamics is governed by the SCE. The partitioning of the energy fraction in spacetime and field dynamics can change in time, but the self- consistency condition between the system (spacetime) and its environment (quantum field) ensures energy conservation.  The Lenz law captures the essence of this type of backreaction on the system by quantum processes of this kind in the environment. Read these two papers for an explicit demonstration of this relation for quantum fields in the homogeneous yet weakly anisotropic Bianchi Type I universe \cite{CalHu87} and for the weakly homogeneous but isotropic FLRW universes \cite{CamVer94}.

Since particle creation originates from the vacuum fluctuations of the quantum field, the balance between the fluctuation energy in the environment and the dissipative energy in the system dynamics can be regarded as a form of {\it fluctuation-dissipation relation}.  We don't have space to delve into how the  noises of quantum fields are defined, as it begins to enter the realm of stochastic gravity \cite{HuVer}, but refer the reader to these two papers \cite{HuSin95,CamVer96} which treated this topic in great detail.


The case for DCE is slightly different, in that there is an external agent which controls how the mirror moves, thus it involves three parties: drive, mirror and field. If the agent's function is to  set the mirror in motion and after the mirror moves at a constant speed stop the drive, then the energy budget will remain unchanged and the mirror-field interplay can be studied under a controlled situation. One can then ask the interesting question, how would particle creation from the quantum field affect the motion of the mirror? In classical physics, a moving object encountering a damping force such as air resistance will gradually slow down, unless, as classical mechanics says, it moves in the vacuum. Here now, allowing for the presence of a quantum field, we ask whether even the vacuum can exert some reactive force on the object. And indeed it does: backreaction of particle creation acts like  the vacuum has some viscous property. This has been calculated for several important quantum field processes in \cite{HuVacVis}. 

One can also ask, in analogy with the backreaction from cosmological particle creation, whether some form of a fluctuation-dissipation relation exists in DCE. A first attempt was made in  \cite{DalvitFDR}\footnote{We feel the treatment of noise therein is incomplete, and are currently working on this problem with S. Butera \cite{DCE-FDR}}. The situation is more complex since there are three parties involved. In the field: if it is in a steady state there is a relation between the causal Green function (the  antisymmetric two point function) responsible for dissipation and the Hadamard function, the symmetric part of the two point function) for the noise. This was the key observation in  \cite{Mottola} for black hole thermal radiance associated with Hawking effect -- notice  the black hole has to be placed in a box to keep it at quasi-equilibrium, otherwise there is no such relation.  There is also a relation between the energy supplied by an external agent to sustain the movement of the mirror or atom   
and the energy associated with the particles produced by the excitation of quantum field fluctuations. The intermediary, and the most interesting party,  is the atom whose internal degree of freedom (idf) interacts with the field and its external degree of freedom (edf) responds to the external agent. How these two interactions play out is the main attraction.  This latter kind of relation is closer to the FDR in the expanding universe cases described above. We shall come back to this topic in the next section after we introduce a new model based on microphysics which can accommodate these three party interactions,  the mirror-oscillator-field (MOF) model.

\section{MOF model for moving atoms or mirrors and  analogous models in cosmology}\label{S:bdker}

In this section we continue to explore suitable   theories or models for a closer and better description of moving mirrors or atoms interacting with a quantum field which can capture the essential quantum field processes, such as particle creation and its backreaction effects. These problems all fall under the new and fast developing field known as quantum opto-mechanics (QOM).  We shall explain the structure of a new theory called the MOF model  and how it can be applied to solving practical problems in QOM.  We then bring up in the spirit of analogue gravity a problem in early universe cosmology which has a similar theoretical structure, namely, preheating the universe after inflation.  The setup of the problem and its solution  may offer some inspiration in how to tackle challenging problems in quantum opto-mechanics. We end with a description of the most recent theoretical developments of the MOF model and  suggest some problems of interest which it can be most usefully applied to. 

The distinct feature of the M(mirror)-O(oscillator)-F(field) model is the oscillator: representation of the internal degrees of freedom in an atom by an oscillator, rather than by two levels, and calling it a `harmonic atom' is not the main feature. The main feature is to endow the mirror with an internal degree of freedom represented by a harmonic oscillator (HO), whose dynamics obeys an equation of motion coupled to those of the quantum field. Making the idf of a mirror  a dynamical variable gives life to what goes on inside the mirror, similar to the micro-physics models of a dielectric plane (e.g., in \cite{HutBar,Buh}) but simpler and more effective (read \cite{LinMOF} about how a single HO, the `atom mirror', can be made to represent a broad-band spectrum).  This route was taken after Galley and one of the present authors found it rather a) restrictive to represent a mirror by  a boundary condition  imposed on the field, usually the Dirichlet condition for a perfect  conductor, because it cannot cope with mirrors which are less than perfect; and b)  difficult when using an auxiliary field as  a constraint, as one needs to deal with the nonlocal aspects in the equations of motion. Structurally the MOF model can be linked to many other operating models of mirror-field systems as described in detail in \cite{GBH}.  The immediate benefit is a good representation of the reflectance or transmissivity of an imperfect mirror \cite{SLH}. But its attractiveness is more in its functionality,  in capturing the primary physics of opto-mechanical systems, which extend from experiments involving atom-mirror-optical systems to LIGO gravitational wave detectors. 
Let us first look at the structure of the MOF model, how it  captures the essence of DCE and DCP (dynamical Casimir-Polder) effects and some generic properties of quantum optomechanical systems. 
Here we follow the description of \cite{SLH} 

\subsection{Structure of the MOF model for Quantum Optomechanics (QOM)}

Consider a point mirror interacting with a massless scalar field in (1+1)-dimensional space-time, the mirror is described by two separate degrees of freedom (dof) represented by harmonic oscillators: i) the  mechanical dof or \textit{mdf}by one oscillator with mass $M$, representing the center of mass (CoM) of a mirror or atom, is suspended in a harmonic potential of frequency $\mho$,  and ii) the internal dof or \textit{idf},  described by another harmonic oscillator of mass $m$ and frequency $\Omega$.  While the \textit{mdf} does not interact with the field directly, the \textit{idf} is bilinearly coupled to the quantum field and constrained to be at the CoM position, meaning, the field takes on values at the location of the atom at a specific time. Thus leads to an indirect interaction between the field and the \textit{mdf}\footnote{The traditional treatment goes by a  radiation pressure type of coupling of the form $Nx$ where $N$ is the number of photons and $x$ denotes the displacement of the mirror.}. The \textit{idf}-field interaction determines all the optical properties of the mirror, as studied in \cite{GBH,SLH}.  

The \textit{idf}-field dynamics, representing the electronic excitations for the case of an atom, has a much faster time scale  compared to those  of the mechanical motion of the atomic center of mass, such that $\Omega\gg\mho$. This is the case for a mirror moving at non-relativistic speed.  
The action in the MOF model is given by
\begin{align}
    S &= \int\! dt\;\left\{ \bkt{\frac{1}{2}M\dot{Z}^2-\frac{1}{2}M\mho^2Z^2}+\bkt{\frac{1}{2}m\dot{q}^2-\frac{1}{2}m\Omega^2q^2} \right. \notag \\ 
    &\qquad\qquad\qquad\qquad+ \left. \int\! dx\; \frac{\epsilon_0}{2}\Bigl[\bkt{\partial_t\Phi^2}-c^2\bkt{\partial_x\Phi}^2+\lambda\dot{q}\Phi\delta(x-Z)\Bigr] \right\}\,,
\label{action}
\end{align}
where we denote the center of mass position  of the \textit{mdf} by $Z(t)$, the amplitude of the \textit{idf} by $q(t)$ and the scalar field by  $\Phi(x,t)$. {Noticing that the free space permittivity in (1+1)-dimensions scales as $\epsilon_0\sim (\text{Charge})^2 \,\text{(Time)}^2\,\text{(Mass)}^{-1}\text{(Length)}^{-1}$, the free field Lagrangian in \cite{GBH} has been rescaled here by a factor of $\epsilon_0$ for dimensional consistency.}.  

For relativistic motion which is required for the treatment of acceleration radiation such as the Unruh effect, one needs to use the proper time,  modify the kinetic terms, and pay attention to the time-slicing scheme. This model will describe a generalization of the Unruh-DeWitt detector theory \cite{Unr76,DeW79,RSG}, an interesting topic which occupies a prominent place in the study of relativistic quantum information theory. See \cite{RHA,HHL,HHLY} for further details.  

In drawing a correspondence between the scalar field used here and an electromagnetic (EM) field, we observe that the free field Lagrangian would correspond to that of an EM field if $\Phi(x,t)$ is to represent the vector potential $A$. We have chosen a form of the bilinear coupling motivated by the electrodynamic form of interaction $\bkt{\sim \dfrac{e}{mc}\,p \cdot A}$, bearing in mind that the mirror's \textit{idf} can also be used to represent the electronic level structure inside a `harmonic' atom. This type of coupling is often referred to as `derivative' coupling. A similar model for describing mirror-field interactions has also been proposed in \cite{WangUnr}. The original MOF model in \cite{GBH} has a different  form of coupling  $\bkt{\sim\lambda q\Phi}$, called `minimal'. 

The $\delta(x-Z)$ factor in the coupling restricts the \textit{idf}-field interaction to the center of mass position and the position dependence of the scalar field in turn leads to an effective force on the \textit{mdf}. We choose the coupling $\lambda$ to have the dimensions of the electronic charge $e$ and $\Phi(x,t)$ to have the dimensions of $A/c$. This is in agreement with the correspondence of the MOF model (see \cite{GBH}) with the Barton-Calogeracos (BC) model \cite{BC}, where in the limit of adiabatic \textit{idf} evolution the coupling $\lambda$ can be physically identified as the surface charge density.


To bring the MOF closer to home,  we show how it can  describe the classical optical and mechanical properties exhibited by a mirror, leading to the known intensity-position radiation pressure coupling. The coupled equations of motion for the classical amplitudes of the \textit{mdf}, \textit{idf} and field 
($\{\bar{Z},\dot{\bar{Z}},\bar{q},\dot{\bar{q}},\bar{\Phi},\dot{\bar{\Phi}}\}$ respectively) are obtained from the action in Eq. (\ref{action}), that ism $\delta S =0$,
\begin{align}
    \ddot{\bar{Z}}(t)+\mho^2\bar{Z}(t) &= \frac{\lambda\,\dot{\bar{q}}(t)}{M}\,\partial_x\bar{\Phi}(\bar{Z},t)\,,\label{class1}\\
    \ddot{\bar{q}}(t)+\Omega^2\bar{q}(t) &= -\frac{\lambda}{m}\,\dot{\bar{\Phi}}(\bar{Z},t)\,,\label{class2}\\
    \epsilon_0\Bigl[\ddot{\bar{\Phi}}(x,t)-c^2\,\partial_x^2\bar{\Phi}(x,t)\Bigr] &= \lambda\,\dot{\bar{q}}(t)\,\delta(x-\bar{Z})\,.\label{class3}
\end{align}
It can be seen that the moving \textit{idf} acts as a point source for the field and the \textit{idf} is in turn driven by $\dot\Phi$ at the center of mass position $\bar{Z}$. 
Also, with $\lambda$ representing the charge density, it can be seen from Eq. (\ref{class2}) that the force on the surface charge degree of freedom goes as $\sim \lambda \dot{\bar\Phi}$.  We have assumed here that the mirror's center of mass velocity is in the non-relativistic limit, such that $\left\vert\dfrac{d\bar{Z}}{dt}\right\vert\ll c$. For a relativistically moving mirror, the \textit{idf} would more generally observe a Doppler shift of the field with respect to the moving center of mass as Eq. (\ref{class2}) becomes
\begin{equation}
    \ddot{\bar{q}}+\Omega^2\bar{q} = -\frac{\lambda}{mc} \left( \dot{\bar{Z}}^0  \partial_t  + \dot{\bar{Z}}^1 \partial_x \right)\bar{\Phi}(\bar{Z}^\mu) \,,
\end{equation}
where $\bar{Z}^\mu = (\bar{Z}^0(\tau), \bar{Z}^1(\tau))$ is the worldline of the mirror parametrized by its proper time $\tau$, and $\dot{O} \equiv dO/d\tau$. As the motion of the mirror's center of mass leads to the motion of the charges at the surface  interacting with the field, the surface charges experience a Doppler shifted field which in turn changes their optical response,  leading to dynamically changing boundary conditions observed by the field.  

Using the MOF model to represent an imperfect mirror's reflectance or transmissivity has been considered in detail in \cite{GBH,SLH}.  With what is described above, it is not difficult to identify the key players -- dynamical variables -- in DCE, DCPE and quantum friction. Once the idf-field interactions and the motion of the mdf (CoM) are stipulated,  one can set up the coupled equations of motion and solve them for the description of their dynamics. The expanse of potential applications of the MOF model is vast and investigations have just begun.


\subsection{An interlude: Cosmological Analogue in Preheating after Inflation}

In keeping with the aims of this paper, i.e., finding the analogies and defining the correspondences between atom/mirror field interactions and quantum field processes in the early universe, we describe one scenario in inflationary cosmology, the  preheating phase after the universe ended inflation and began to enter the reheating phase, eventually warmed up to become the radiation-dominated FLRW universe we are at home with.  This case study intends to serve two  purposes relevant to the themes we have discussed so far: 

A) {\it Backreaction is important}. We have discussed the importance of the backreaction of particle creation at the Planck scale on the dynamics of the background spacetime.   
Backreaction effects of particles produced from the parametric excitation of the fluctuations of the inflaton field at the GUT scale is also of primary importance. Their backreaction on the background inflaton field is responsible for preheating the universe and eventually damping it out, driving the universe to its true vacuum. 

B) {\it tripartite interaction}. The three key players we talked about in QOM have their analogues here, except for the relative importance of the roles they play. The emphasis in QOM is more on the atom or the mirror:  how its external or mechanical dof changes the field in the case of moving atom or mirror  and how the field affects its idf, whereas in cosmology the attention is more on the field and its fluctuations.    

1) classical background spacetime with metric $g_{mn}$ , like the  mechanical dof $Z$ for the CoM motion of the mirror or atom;
 
2) inflaton quantum field $\Phi (t) = \phi (t) + \tilde \phi (t)$ where $\phi$ is the background inflaton field, and $\tilde \phi (t)$ its  fluctuations.  

3) the fluctuations $\tilde \phi$ of inflaton field acts like the idf of a mirror or atom

The dynamics proceeds in several stages: 
Notice first that
1) always enters in 2) and 3), but the  backreaction effects of 2) and 3) on the background spacetime geometry are insignificant, since the GUT energy scale is five orders of magnitude lower than the Planck scale. The main stage is on how particle creation from the parametric amplification of the quantum fluctuations of the inflaton field changes the dynamics of the inflaton field,  which in turn changes the cause of the universe.

At the end of slow-roll (quasi-de Sitter) phase, the inflaton field starts falling into a steep potential well, that is the beginning of the (pre)heating phase. The rapid oscillation of 2) in this phase parametrically amplifies 3) $\tilde \phi$ , ample particle creation ensues. The backreaction of 3) on 2) causes $\phi (t)$  to damp away quickly, eventually settles the inflaton field $\Phi (t)$ to the minimum of the potential well. This particle creation process based on the same parametric amplification mechanism -- like that at the Planck scale being responsible for the isotropization and homogenization of the universe  -- is responsible for the preheating the universe and the eventual ushering-in of the radiation-dominated FLRW phase.   

Thus, preheating after the inflationary stage serves as an illustration of this tripartite dynamical variable scheme analogous to the MOF model. For a thorough treatment using QFTCST and SCG methodologies, see \cite{RamHu97}. 

\subsection{Developments and applications of the MOF theoretical framework}

Returning to the optomechanical side of the analogue we wish to end by pointing out some current work   furthering the theoretical developments of MOF model and identify interesting problems this model can be applied to finding meaningful solutions. 

i) Not long after the first two papers on MOF \cite{GBH,SLH} appeared, Lin \cite{LinMOF} continued to explore the features of MOF model in relation to the functionality of the Unruh-DeWitt detector, which in (1+1) dimension he referred to as an atom mirror. Lin showed that when coupling between the oscillator representing the mirror's idf  and the quantum field is strong, a broad frequency range of the quantum field can be mostly reflected by the atom mirror at late times. Constructing a cavity model with two such atom mirrors, he demonstrated  how the
quantum field inside the cavity evolves from a continuous to a quasi-discrete spectrum which gives a
negative Casimir energy density at late times. 

The differences between the two types of coupling with regard to the reflectivity properties of an imperfect mirror as a function of the normal mode  frequency $\omega$ of the field in ratio to the idf oscillator frequency $\Omega_0$ are clearly shown in Figure 1 of his paper, based on their late time behavior. In the static limit $\omega \to 0$    
the minimal-coupled  and the derivative-coupled detectors act like a dielectric and a metal mirror respectively, as noted earlier in \cite{GBH}  for minimal-coupling and in \cite{SLH} for derivative-coupling.

Another reassuring find is that the late-time renormalized energy density of
the field inside a cavity of the atom  mirrors is always negative even in the weak oscillator-field coupling regime where the cavity modes are few.  

ii) Sinha, Lopez \& Suba{\c s}{\i} (2021) \cite{SinLopSub} used the MOF theoretical framework and studied the {\it nonequilibrium dissipative dynamics of the center of mass (CoM) of a particle } coupled to a field via its internal degrees of freedom.  Notice even the field is directly coupled only to the idf of the atom or mirror,  the field value is assumed at the spacetime point of CoM. Therefore there is an indirect nonlinear coupling between the field and the CoM.   They used the influence functional approach to account for the self-consistent backaction of the different degrees of freedom on each other, including the coupled nonequilibrium dynamics of the internal degree of freedom and the field, and their influence on the dissipation and noise of the center of mass.   they  assumed a weakly nonlinear coupling and adopted a functional perturbative method \cite{HPZ93} to derive the influence action and from there obtain the effective equation of motion for the CoM of the atom or mirror.  

iii) {\it motional decoherence}: An interesting physics which exemplifies the indirect coupling between the CoM of an atom and its idf via the direct coupling between the atom's idf and the field is motional decoherence.  Movement of the atom's CoM changes the  field, and the disturbance is registered in the idf of atom, causing dephasing or decoherence. Dalvit and Neto treated this problem by way of radiation pressure \cite{DalNet}.   It would be pleasing to see this redone via the MOF model, a prototype of which was studied by Shresta and Hu \cite{ShrHu}. The nonequilibrium dynamics  formalism of Sinha et al \cite{SinLopSub} is expected to provide a more thorough analysis  of this problem. 

iv) Butera (2022, 2023). In \cite{But2Mir} Butera used the MOF theoretical framework to provide a microphysics upgrade of the traditional radiation pressure coupling between a mirror's mechanical dof and the field \cite{Law,BC}. Owing to the nonlinear nature of their coupling, Butera adopted the perturbative functional approach,  same as  in Sinha, et al \cite{SinLopSub} to derive a second-order effective action for the system of  two mirrors. He found that the quantum and thermal fluctuations of the field appear in the form of colored noises acting on the mirrors, whose nonequilibrium dynamics is non-Markovian. He also derived the quantum Langevin equations and showed the existence of  a generalized fluctuation - dissipation relation. In \cite{ButN-D}, using second-order perturbation theory, he derived the master
equation governing the mechanical motion of the mirror. His analysis reveals that the mirror experiences coloured noise and non-local dissipation, which originate from the emission of
particle pairs via the dynamical Casimir effect.   

v) {\it Graded influence action.}  The MOF model is a simplified version of a grander theoretical framework known as graded influence action, which refers to a nested set of coarse-grained effective actions \cite{cgea} 
where one can deal with the interactions of many dynamical variables in an orderly manner. Graded refers to an ordering of which variables are to be coarse-grained first, forming the first tier of effectively open system, followed by another variable, forming the second tier of effectively open system, and so on. At the end, one would come up with the sum total effect of all the variables with direct or indirect interaction with the one variable of special interest to us, by way of  a nested set of influence actions. A vivid demonstration of the application of this formalism is shown in the derivation of the atom-dielectric forces by Behunin and Hu in \cite{BehHu10} and \cite{BehHu11} (see e.g., Figure 1 for the orderly coarse-graining of five variables)  which the reader is referred to for greater details and insights. A work in the same vein investigating the effects of a medium-altered quantum field on an atom is currently in preparation \cite{HHGIA}.   

vi) {\it Quantum friction}  A neutral atom moving at uniform speed slight above a dielectric plate will experience a drag force, which is commonly known as quantum friction\footnote{As remarked in our earlier discussion of analogy with the backreaction due to particle creation, a better term would be {\it quantum viscosity}, as friction connotes a simple resistive force but in reality this is a reactive force, with phase dependence and memories, like that in an inductor.}. For theory work which resonate with our emphasis and concerns, see, e.g., \cite{IntBehQF}. This phenomenon is a good arena to see how a graded influence action formalism is played out, similar to the way Behunin and Hu derived the atom-dielectric force: First calculate the field modified by a medium, then calculate how the idf of the moving atom interacts with the medium-altered field. Noticing that the field value is taken at the spatial point  where the CoM of the atom is located at that instant, that's where the movement of the atom enters into the picture. The backreaction of the medium-altered quantum field on the CoM of the atom is what gives the drag force known as quantum friction. Similar to the backreaction of particle creation on the dynamics of the universe, there should also be a fluctuation-dissipation relation depicting the balance between the moving atom and the fluctuations in the quantum field. And,  if the atom's motion is kept at a constant speed by an external agent, the agent should experience a reactive force from the atom, which when traced back, originates from the medium-altered quantum field fluctuations. This nested set of influences captured by the graded influence action is what makes this effect so interesting and this method so empowering.

\quad\\
\noindent{\bf Acknowledgment}  B.-L. Hu thanks his co-authors S. Butera and Y. Xie in \cite{XBH} for discussions on points made in Sec.~\ref{S:eoet}, and his co-authors R. Behunin in \cite{BehHu10,BehHu11} for the establishment  of the graded influence action and C. Galley, S.-Y. Lin and K. Sinha in~\cite{GBH,SLH} for the construction of the MOF model presented in Sec.~\ref{S:bdker}. An anonymous referee of \cite{XBH} is also to be thanked for his/her asking us many pertinent questions on topics discussed in Sec.~\ref{S:eoet} for better clarification, thus enticing us to appreciate a closer to reality perspective. J.-T. Hsiang is supported by the National Science and Technology Council of Taiwan, R.O.C. under Grant No.~NSTC 113-2112-M-011-001-MY3. B.-L. Hu enjoyed the warm hospitality of colleagues at the Institute of Physics, Academia Sinica, Taiwan, R.O.C. where part of this work was done.  


\end{document}